# Conceptualization, Operationalization, and Measurement of Machine Companionship: A Scoping Review

Jaime Banks
School of Information Studies
Syracuse University
Syracuse, NY, USA
banks@syr.edu

Zhixin Li
School of Information Studies
Syracuse University
Syracuse, NY, USA
zli457@syr.edu

## ABSTRACT

The notion of machine companions has long been embedded in social-technological imaginaries. Recent advances in AI have moved those media musings into believable sociality manifested in interfaces, robotic bodies, and devices. Those machines are often referred to colloquially as "companions" yet there is little careful engagement of machine companion*ship* (MC) as a formal concept or measured variable. This PRISMA-guided scoping review systematically samples, surveys, and synthesizes current scholarly works on MC ($N$ = 71; 2017-2025). Works varied widely in considerations of MC according to guiding theories, dimensions of *a-priori* specified properties (subjectively positive, sustained over time, co-active, autotelic), and in measured concepts (with more than 50 distinct measured variables). We ultimately offer a literature-guided definition of MC as an autotelic, coordinated connection between human and machine that unfolds over time and is subjectively positive.

## 1 Introduction

Fictional worlds abound with artificial intelligences—Bender, Bubo, Bumblebee, Data, Dot Matrix, Frankenstein, GlaDOS, JARVIS, Johnny 5, Kitt, Maria, Marvin, Max, R2D2, Robby, Samantha, Sonny, Talos, Tin Man—engaging humans as friends, cohabitants, guardians, and even lovers. Human-machine relationships have generally been relegated to hopes and fears embedded in those fictions, since robots and non-embodied agents were functionally limited and wholly not convincing in their sociality. Recent advances in large-language models (LLMs) and their natural-language interfaces, however, have moved those fictional affairs into actuality. In addition to long-standing connections people form with cars, game avatars, and toys, some machines now have capacities to communicate autonomously such that the potential for operationally and phenomenologically real relationships must be considered.

As yet, however, there is little systematic attention to the ways machines may meaningfully function as *companions*—as sustained, intimate affiliates. This work begins to address that gap through a PRISMA-guided scoping review of current literature on close human-machine relationships, synthesizing theoretical and empirical work to offer a conceptual definition of machine companionship (MC), explore its operational dimensions, and catalog current measurement approaches.

### 1.1 Companionship Among Humans

The term "companion" is used rather casually in everyday language, and even in scholarly literature, to refer to something that exists alongside another thing—in that sense, a handbook is a companion in learning, a utensil is a companion in cooking, and a camera is a companion in travel. Through the lens of close interpersonal relationships, though, companions are familiar individuals—like family, friends, peers, neighbors—with whom we enjoy spending time and engaging in positive and pleasurable activities (Sarason & Sarason, 2001).

Companion*ship*, then, can be understood as a relational process of positive, pleasurable social exchange (Rook, 1987) or as a state characterized by the reliable presence of such exchange. As part of close personal relationships, companionship can evolve longitudinally—incrementally maintained or dissolved over time (see Troutman & Fletcher, 2010), allowing for exchanges to be anticipated and normed (Hall, 2012). It is co-active in that participants in the relationship engage one another in shared activities (Lüscher et al., 2022). Its dual nature as both process and state permits the empirical examination of both its operation (how the process unfolds) and its construction (through individual, joint, and social construals; see Dreher, 2009). Companionship is positive or pleasurable, such that co-engaged parties enjoy the shared activities (Rook, 1987) and, because of that enjoyment, companionship is autotelic—it is engaged for its own sake, as an end in itself (Wilks, 2010). Of note, other definitions of companionship entangle the concept with the provision of social support. However, we argue in alignment with Rook (1987) that need satisfaction, including needs for social support, is an *outcome* of companionship. This is not to say that companionship does not satisfy needs or is not driven by individual desires for need satisfaction (e.g., Deci & Ryan, 2014). Rather, understanding companionship requires parceling it out from its effects (e.g., wellbeing; as in Alsarrani et al., 2022).

A core feature of human companionship is intimacy (Baron et al., 2009) built through mutual self-disclosures (Altman & Taylor, 1973) and, sometimes, physical closeness (Ganguli, 2024). Companionship may also involve attachment, characterized by emotional closeness and a sense of security, as well as social integration—a sense of belonging through shared interests, concerns, and activities (Weiss, 1974). Friends, romantic partners, colleagues, family members, and cohabitants can all participate in companionship relations, so long as the association is maintained through intrinsically pleasurable activities. Notably, companionship in a relationship could be one-sided—one may be said to provide companionship to another, while the other doesn't necessarily offer the same to the first (see Stein et al., 2024).

## 1.2 Considering Machines as Companionship Participants

We have, above, conceptually characterized companionship as an autotelic connection in which two participants engage in sustained, co-active, and pleasurable activities. This conceptualization focuses on relational functions rather than being anchored to the performance of human abilities (see Banks & de Graaf, 2020) such that both humans and machines can participate in companionship relations. We here consider the possibilities for machines to be companionable, taking up the requirements for autotelic, sustained, co-active, pleasurable activities, myriad machines could engage in companionship relations. Piloted vehicles, mobile phones, and avatars have all been considered as companions despite their non-autonomy (e.g., Xiao, 2020) and (semi-)autonomous machines like vehicle-embedded characters and LLM-based personas (e.g., Brandtzaeg et al., 2022) are candidate companions given their apparent social agency.

We acknowledge that, outside our purposefully ontology-agnostic definition, some argue for necessary conditions for a machine to be a companion: They must have intentionality, predictable behavior, reasoning and planning competencies, ability to anticipate human behavior (i.e., theory of mind), require little effort, and understand a human as an individual (Pulman, 2007). However, some of those criteria are often not even required of human companions—human relationships are often incidentally or purposefully effortful and we regularly have breakdowns in reasoning and behavioral prediction (e.g., Kelly et al., 2017). Alternately, relinquishing anthropocentric views on intelligence could advance machine companionship different from anything we might imagine today (Taylor et al., 2010).

Potentials and perils of machine companionship are increasingly considered proximal, rather than being far-fetched issues linked with artificial general intelligence or the ostensible singularity. There may be benefits gleaned from machine companionship: Customization of designed companions could lead to greater satisfaction than in human companionship (Levy, 2007), along with improvements to social health and diminution of negative experiences like rejection, facilitating social interactions with others, and modeling prosocial behaviors (see Prescott & Robillard, 2021). Alongside these optimistic accounts are critiques of the authenticity and ethics of such relationships. Scholars have argued machines lack subjectivity, reflexivity, or interestedness—the foundations of personhood—so they cannot participate in authentic friendships (Archer, 2021). Some go further, suggesting that machines, if used as relational partners, must remain clearly subordinate, lest their treatment distort human values or dehumanize their users (Bryson, 2010). Such critiques altogether reflect what Turkle (2007, p. 502) has called a "crisis of authenticity" in human-machine relations. In line, emerging perspectives encourage attention to and safeguarding against risks in these relationships: As intimacy emerges there may be increased privacy and safety hazards, exposure to harmful content, dependency that links users to corporate interests, echo-chamber effects, and spillover to human relationships (Starke et al., 2024). Others still argue that any shortcomings are merely technical shortcomings that may be overcome with time (Danaher, 2019).

## 1.3 Objectives of this Review

Extant literature includes systematic reviews of machine companionship through varied lenses. A review by Pentina and colleagues (2023) examined consumer-machine relations finding primarily quantitative, cross-sectional work across social psychology, communication/media, and human-machine interaction. Rogge (2023) covered a 22-year span to consider scholarly engagement with artificial companions, identifying defining qualities of adaptivity and engagement, along with design properties of user adaptivity, context adaptivity, engagement behaviors, personality, and appearance. Another considers companion robot morphologies (anthropo- and zoo-morphic favored over more functional machines) and inattention to the role of robot mobility and co-operation (Ahmed et al., 2024). Chaturvedi and colleagues (2023) found heightened attention to personification and embodiment, conversation and interaction, loneliness mitigation, anthropomorphism and social closeness, and AI as distinctly social. Among other scoping reviews, many focus only on the application of machines for health outcomes (e.g., loneliness reduction; Gasteiger et al., 2021). These works each attend to patterns in scientific efforts, in publication or citation trends, in the machine design or operation; some narrowly considered robots, while others excluded robots or chatbots or voice assistants. Many of those studies also propose conceptual models of antecedents, mediators, and outcomes alongside user and situation characteristics.

Our aim in this review is different. Despite the centrality of companionship to how human-machine relations are considered in technical development, social psychology, news and fictional narratives, and public imaginaries, there is yet no systematic effort to synthesize relevant works toward a consensus on the basic construct—the conceptual or operational definitions of machine companionship, or on its dimensionality and measurement. It is also critical to synthesize theoretical work alongside the empirical, since emerging critiques often come in advance of our ability to observe actual human-machine interactions; technology lags behind imaginaries. A scoping review is an appropriate approach for addressing these high-level questions across social-scientific fields relevant to the notion of companionship.

To begin closing that knowledge gap, this inquiry addresses the following questions in relation to current literature:
**RQ1**: How is machine companionship conceptually defined?
**RQ2**: How is machine companionship operationally defined?
**RQ3**: How is machine companionship measured?

## 2 Method

This scoping review followed an adaptation of PRISMA guidelines for qualitative scoping reviews (i.e., PRISMA-ScR; Tricco et al., 2018). This framework ensures that comprehensive reviews are rule-based (minimizing selection biases) and fully reported (ensuring transparency and replicability), ensuring readers can assess the scope and validity of findings. Data processing was facilitated by the Covidence (covidence.org) systematic review platform. All materials are available in the online supplements for this project: https://osf.io/hgq54/

### 2.1 Information Sources

Databases were selected based on their likely coverage of core concepts of social machines and/or companionship. Two broad databases, Scopus and Web of Science, were chosen for their inclusion of both technology and social psychology topics. EBSCO (inclusive of PsycInfo, PsycArticles, PsycTESTS), was selected to capture literature within social psychology. Two databases—IEEE Xplore and ACM Digital Library—were identified as likely hosts to MC-related literature from more technology-oriented and interdisciplinary venues. Our preliminary literature review indicated highly relevant works were hosted in the arXiv preprint repository. Although those works have often not yet undergone peer review, many had already been heavily cited, reflecting impact on current MC theorizing that also warranted review.

### 2.2 Eligibility Criteria

Eligibility criteria were guided by the PICOS (Population, Intervention, Comparator, Outcome, and Study Design) framework (Methley et al., 2014). Works were initially screened for inclusion criteria (Table 1). We additionally (a) required MC to be the central topic of the paper and (b) excluded other systematic reviews. Works were also screened for adherence to the original search criteria: English-language, published 2017-2025, and peer-reviewed journal articles or proceedings or preprint equivalents thereof. Book chapters, workshops, editorials, and other non-peer-reviewed formats were excluded.

### 2.3 Search

In line with our tentative definition for companionship, the literature search strategy emphasized identifying literature that conceptually or empirically engages the state of a social machine being in some relational association with humans. Specifically, we targeted works that consider a relational association (a) over time, (b) interactive, (c) ranging in depth from familiarity or friendship to intimate or romantic, but always subjectively positive, and (c) engaging in this relation for its own sake or social value, rather than toward applied or practical outcomes.

The search strategy, developed by the researchers in consultation with a university research librarian, leveraged the two conceptual components of the construct of interest—machines and companionship. A set of search terms was developed for each. To generate the terms, Author 2 (A2) completed an initial exploratory search of recent literature across major academic databases using topical keywords (e.g., virtual companion, social AI, AI companion*) and identified and catalogued prevalent terms across works. We reviewed the candidates, removed terms that are theoretically adjacent to but not inherent to our working definition of companionship as a sustained, co-active, autotelic affiliation (e.g., "trust" as a potential element of but alone not sufficient to constitute companionship).

Table 1
PICOS-based Eligibility Criteria Matrix

| PICOS Criterion | Requirement – Empirical Works | Requirement – Theoretical Works |
|---|---|---|
| Population | Adults who interact with MCs | None |
| Intervention | Interaction with primarily social MCs (i.e., not merely practical/applied) on at least three occasions; *or* data accounting for repeated interactions | Argumentation attends to primarily social MCs (i.e., not merely practical/applied) |
| Comparator | None (all inclusive) | None (all inclusive) |
| Outcome | Subjective or relational experiences (i.e., not only task-based, productive, or practical outcomes) | None (all inclusive) |
| Study Design | Any empirical method/tradition, excepting technical or design works | Any theoretical method/tradition, excepting technical or design works |

The resulting search logic was: ("Machine" OR "AI" OR "Artificial Intelligence" OR "social robot"" OR social bot" OR "chatbot" OR "conversational AI" OR "generative AI" OR "conversational agent" OR "voice assistant" OR "virtual") AND ("Companion" OR "companionship" OR "romance" OR "friend*" OR "relationship" OR "intima*" OR "love" OR "attachment").

The search was performed between October 3 and October 9, 2024; it was constrained to English-language works published between 2017 (the year the first mainstreamed AI companion, *Replika*, was released) and 2025 (i.e., inclusive of advance online publications). Understanding there was attention to machine companionship prior to this (e.g., Taylor et al., 2010), we elected to focus on contemporary works considered in relation to current AI abilities and potentials. The search included peer-reviewed journal articles and proceedings for the formal databases and was limited to titles and keywords; arXiv search was based on title and abstract only (as it offered no keyword search). The search was implemented differently across databases, based on the affordance of the search interface. See supplements for database-specific procedures.

### 2.4 Source Screening and Selection

The search produced an initial catalog of 5,708 works which were uploaded to Covidence; the platform automatically de-duplicated and removed 706 entries. Two more duplicates were manually identified, resulting in a total of 5,002 works advanced to screening. The first and second authors (A1 and A2) used

Covidence to independently screen the title, abstract, keywords, and publication type for adherence to inclusion criteria. Each indicated for a given work: "Yes," "no," or "maybe." Where the "yes" or "no" decisions aligned, those works were piped into lists for included and excluded works, respectively. Where decisions diverged or at least one decision was "maybe," works were piped into a conflict-resolution list *(n* = 168). The authors discussed and resolved disagreements for those works. This resulted in the exclusion of 4,875 works and 127 sources for further screening. In a second phase, A2 conducted a more careful reading of full texts to ensure alignment with all PICOS criteria; where exclusion was considered, A1 also conducted a full-text review. A1 and A2 discussed and came to consensus on inclusion/exclusion. In this phase, an additional 56 works were removed, resulting in a final corpus of 71 sources (see Fig. 1).

**Figure 1**
PRISMA-guided Sampling, Screening, and Data-Charting Workflow

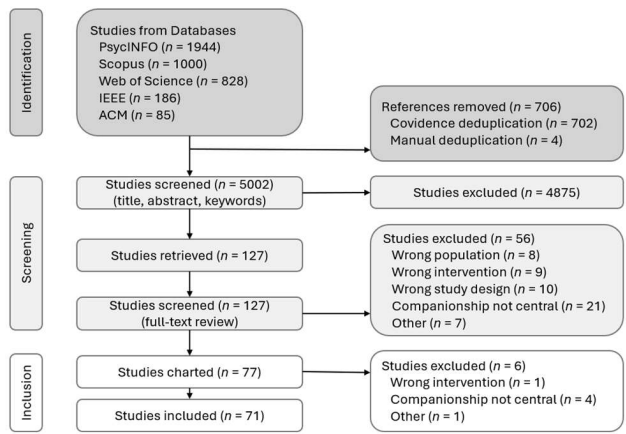

### 2.5  Data Charting Process and Items

Data was fully charted for the final ($N$ = 71) sources. A2 performed the initial charting of each paper, identifying content aligned with predefined extraction categories. For all works, extracted data included general information (e.g., publication metadata, class of paper, and focal machine type), theoretical frameworks, and conceptual and operational definitions of MC. In instances where explicit definitions or conceptualizations were absent but closely adjacent concepts were present (e.g., love, friendship, intimacy), those characterizations were extracted instead. For empirical works, the dimensions measured were also charted. Additional features were extracted but are not analyzed here; see supplements for the complete charting details.

To verify reliability of extracted data after A2 completed initial extractions, A1 independently extracted data from a randomly selected 10% subset (8 papers). Comparisons between A1 and A2 extractions were generally aligned, although variations emerged for some conceptual definitions and companionship dimensions. For machine companionship conceptualization, discussion indicated A1 had taken an approach more attentive to extracting definitions of companionship-adjacent concepts while A2 was more attentive to characterizations of companionship-affording machines. For companionship dimensions, A1 had attended more narrowly to explicit and measured dimensions while A2 was more inclusive and extracted even briefly mentioned dimensions. To reconcile these two divergent extraction approaches, A2 revisited the entire dataset, re-extracting conceptual definitions and dimensions to ensure comprehensive consistency and integrated these extractions with A1's data to produce a finalized dataset. A final comprehensive review by A2 confirmed consistency and coherence throughout the dataset, with any lingering discrepancies merged and documented.

During discussions, six additional works for exclusion based on deviation from inclusion criteria: The focal machine being a human-manipulated avatar ($n$ = 1), companionship/relationality not being the focus of the work ($n$ = 4), emphasis on functional rather than autotelic relations ($n$ = 1). See supplements for the complete charted data set, including resolutions.

### 2.6  Synthesis of Extracted Data

A1 performed inductive thematic analysis (Braun & Clarke, 2006) for each extracted variable, separately. This entailed a deep reading of extractions, assignment of one or more open codes to the extracted data, iterative collapsing of codes into categories and categories into themes. In the post-reduction naming of themes, extant literature was engaged to consider how patterns align with or diverge from known companionship dynamics. That literature engagement followed analysis; although our work was sensitized by *a priori* parameters for the investigation (see Bowen, 2019), we worked toward first allowing the data to speak for itself. Following the reduction and naming, A2 conducted a face-validity check by reviewing the thematic hierarchy in relation to the documented analysis process. No validity issues were identified.

The specific analyses undertaken include those for general data-set descriptives (meta-data, scholarly tradition and field, machine types, methods engaged, and sample attributes). RQ1 was addressed through synthesis of the conceptual definition and theoretical framework, RQ2 through the dimensions identified in relation to the working definition of MC, and RQ3 by cataloging measurement approaches (empirical only). See supplements for code-reduction documentation for this analytical process.

## 3  Results

### 3.1  Characteristics of Sources

The sources in our review ($n$ = 71) were categorized along four descriptive dimensions: Class of paper, tradition, machine type discussed, and first-author institution country (Table 2, Figure 2). Most works were empirical (i.e., original data collection or secondary analysis; $n$ = 45). Of these, 26 recruited participants with prior experience with the focal machine (e.g., voice assistant owners); three focused solely on first-time users, two included both experienced and novice groups, and 14 did not report prior familiarity. Sample sizes varied widely ($N_{range}$ 3–1,381) and recruitment relied on convenience sampling (e.g., online panels, existing user communities). Most others were theoretical ($n = 23$), engaging in conceptual analyses or argumentation, including work centered on theories and concepts, ethics, or speculation. Three papers otherwise reported interpretive work through psychoanalysis, literary theory, and media criticism.

The most prevalent tradition comprised social-scientific works ($n = 49$) grounded in communication, psychology, and human-computer interaction. Philosophical traditions were present in 11 papers, focusing on ontological, epistemological, and conceptual concerns. Ethical approaches appeared in five papers, analyzing topics like emotional deception, value alignment, consent, and design responsibility. The remaining six papers drew from critical traditions, including feminist bioethics, posthumanism, and Foucauldian analysis.

Works considered various artificial agents (Figure 2). Most common were robots ($n = 26$), particularly humanoid or socially assistive robots; voice assistants (e.g., Amazon's Alexa; $n = 21$) were also common, often in the context of domestic routines or emotional interaction. Text-based chatbots (e.g., Mitsuku) and AI companions (e.g., Replika) were prominent in discussions of relational dynamics, friendship, and affective attachment. Other machines were digital characters (both human and non-human forms), holograms, and LLMs. In eight cases, more than one machine type was considered.

Lastly, geographic spread of scholarship was assessed according to the institution affiliation of the first author. A plurality was associated with the United States ($n = 19$), followed by contributions from China ($n = 9$), the United Kingdom ($n = 5$), and Norway ($n = 5$). Additional regions represented included East Asia, Europe, Middle East, and Latin America. A detailed account of these distributions is provided in Table 2. Throughout the results, references are made to sources by indicating the ID number from Table 2 in brackets (e.g., [1]).

**Figure 2**

Discussed Machine Companions in Reviewed Studies by Country and Publication Year

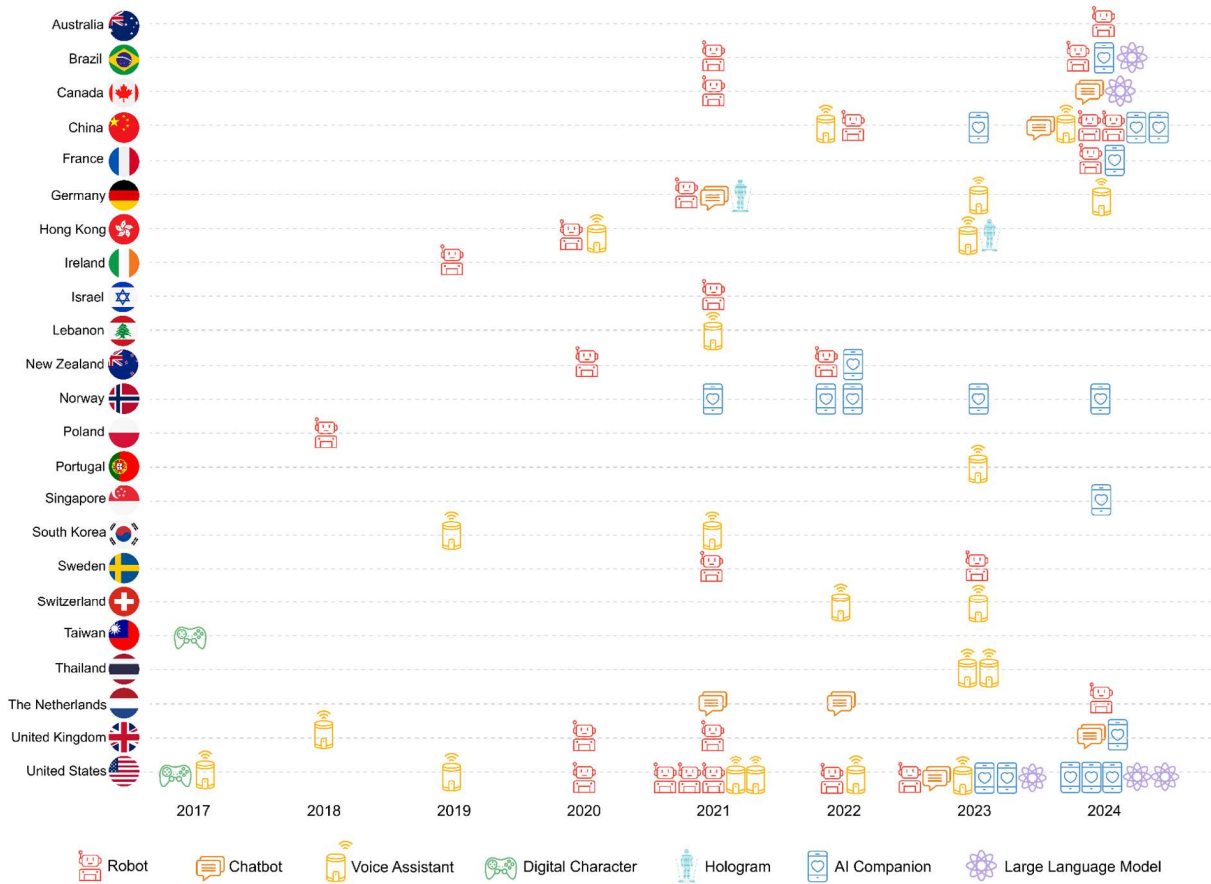

*Note.* This figure maps the diversity of machines discussed in the reviewed studies. Symbols indicate the type of machine discussed, organized by country (y-axis) and final publication year (x-axis). The order of machine types at the bottom reflects the approximate chronology of their emergence in public and research domains, with robots and chatbots appearing earlier, and AI companions and LLMs emerging most recently. Advance-online works with 2025 dates are grouped here with 2024 works.

Table 2

Characteristics of Reviewed Studies by Class, Primary Tradition, and First-Author Institution Country

| ID | Paper | Class | Primary Tradition | Institution Country |
|---|---|---|---|---|
| 1 | Banks (2017) [online 2015] | Empirical | Social Scientific | United States |
| 2 | Aibel (2017) [online 2016] | Psychoanalytic | Critical | United States |
| 3 | Li et al. (2018) [online 2017] | Empirical | Social Scientific | Taiwan |
| 4 | Dziergwa et al. (2018) [online 2017] | Empirical | Social Scientific | Poland |
| 5 | Alexander & Yescavage (2018) | Film/Television | Critical | United Kingdom |
| 6 | Danaher (2019) | Theoretical | Philosophical | Ireland |
| 7 | Caudwell & Lacey (2019) | Theoretical | Critical | New Zealand |
| 8 | Cho et al. (2019) | Empirical | Social Scientific | South Korea |
| 9 | Pradhan et al. (2019) | Empirical | Social Scientific | United States |
| 10 | van Maris et al. (2020) | Empirical | Social Scientific | United Kingdom |
| 11 | Ki et al. (2020) | Empirical | Social Scientific | Hong Kong |
| 12 | Huang et al. (2020) | Theoretical | Social Scientific | United States |
| 13 | Fröding & Peterson (2021) [online 2020] | Theoretical | Ethical | Sweden |
| 14 | Croes & Antheunis (2020) | Empirical | Social Scientific | The Netherlands |
| 15 | Dubé & Anctil (2021) [online 2020] | Theoretical | Philosophical | Canada |
| 16 | Joshi et al. (2020) | Empirical | Social Scientific | United States |
| 17 | Ramadan et al. (2020) | Empirical | Social Scientific | Lebanon |
| 18 | Li & Sung (2021) | Empirical | Social Scientific | South Korea |
| 19 | Skjuve et al. (2021) | Empirical | Social Scientific | Norway |
| 20 | Hoorn (2020) [online 2021] | Theoretical | Social Scientific | Hong Kong |
| 21 | Ryland (2021) | Theoretical | Philosophical | United Kingdom |
| 22 | Erel et al. (2021). | Empirical | Social Scientific | Israel |
| 23 | Fox & Gambino (2021) | Theoretical | Social Scientific | United States |
| 24 | Weber-Guskar (2021) | Theoretical | Ethical | Germany |
| 25 | Kim & Choudhury (2021) | Empirical | Social Scientific | United States |
| 26 | Carvalho-Nascimento et al. (2021) | Theoretical | Ethical | Brazil |
| 27 | Shao & Kwon (2021) | Empirical | Social Scientific | United States |
| 28 | Setman (2021) | Theoretical | Ethical | United States |
| 29 | Song et al. (2022) | Empirical | Social Scientific | China |
| 30 | Uysal et al. (2022) | Empirical | Social Scientific | Switzerland |
| 31 | Brandtzaeg et al. (2022) | Empirical | Social Scientific | Norway |
| 32 | Zhang (2022) | Theoretical | Philosophical | China |
| 33 | Croes et al. (2022) | Empirical | Social Scientific | The Netherlands |
| 34 | Reineke (2022) | Theoretical | Philosophical | United States |
| 35 | Xu & Li (2022) | Empirical | Social Scientific | United States |
| 36 | Skjuve et al. (2022) | Empirical | Social Scientific | Norway |
| 37 | Healy (2024) [online 2022] | Theoretical | Philosophical | Australia |
| 38 | Sheng & Wang (2022) | Literary | Critical | China |
| 39 | Weijers & Munn (2022) | Theoretical | Philosophical | New Zealand |
| 40 | Pentina et al. (2023) [online 2022] | Empirical | Social Scientific | United States |
| 41 | Ali et al. (2023) | Empirical | Social Scientific | China |
| 42 | Skjuve et al. (2023) | Empirical | Social Scientific | Norway |
| 43 | Guerreiro & Loureiro (2023) | Empirical | Social Scientific | Portugal |
| 44 | Wienrich et al. (2023) | Empirical | Social Scientific | Germany |
| 45 | Pal et al. (2023) | Empirical | Social Scientific | Thailand |
| 46 | Tschopp et al. (2023) | Empirical | Social Scientific | Switzerland |
| 47 | Pal et al. (2023) | Empirical | Social Scientific | Thailand |
| 48 | Dehnert & Gunkel (2025) [online 2023] | Theoretical | Critical | United States |
| 49 | Pan et al. (2024) [online 2023] | Empirical | Social Scientific | China |
| 50 | Marriott & Pitardi (2023) | Empirical | Social Scientific | United Kingdom |
| 51 | Yeung et al. (2024) [online 2023] | Empirical | Social Scientific | Hong Kong |
| 52 | Khan (2023) | Theoretical | Social Scientific | Sweden |
| 53 | Leo-Liu (2023) | Empirical | Social Scientific | Hong Kong |
| 54 | Brinkman & Grudin (2023) | Empirical | Social Scientific | United States |
| 55 | Xygkou et al. (2024) | Empirical | Social Scientific | United Kingdom |
| 56 | Ma & Huo (2025) [online 2024] | Empirical | Social Scientific | China |
| 57 | Xu et al. (2024) | Empirical | Social Scientific | China |
| 58 | Puzio (2024) | Theoretical | Philosophical | The Netherlands |
| 59 | Barbosa & Costa (2024) | Theoretical | Philosophical | Brazil |
| 60 | Guingrich & Graziano (2025) [online 2024] | Empirical | Social Scientific | United States |
| 61 | Mlonyeni (2025) [online 2024] | Theoretical | Ethical | Norway |
| 62 | De Freitas et al. (2024) | Empirical | Social Scientific | United States |
| 63 | Root (2025) [online 2024] | Theoretical | Philosophical | France |
| 64 | Nash (2024) | Theoretical | Critical | United States |
| 65 | Sarigul et al. (2024) | Empirical | Social Scientific | Germany |
| 66 | Banks (2024) | Empirical | Social Scientific | United States |
| 67 | Andreescu (2024) | Theoretical | Philosophical | United States |
| 68 | Li & Zhang (2024) | Empirical | Social Scientific | Singapore |
| 69 | Fan et al. (2025) [online 2024] | Empirical | Social Scientific | China |
| 70 | Pan & Mou (2024) | Empirical | Social Scientific | China |
| 71 | Nikghalb & Cheng (2024) | Empirical | Social Scientific | Canada |

## 3.2 Conceptualizations of Machine Companionship

Among analyzed works, 31 explicitly defined companionship or an adjacent sampling keyword (e.g., friendship, attachment, bond). Four [3, 27, 56, 67] explicitly named and defined the focal relations as companionship. Others defined friendship [6, 31, 39, 59] or relationships generally [2, 7, 15, 19, 20, 23, 33, 38, 64, 65], ecorelationality as an associative ethic [58], affective relationships [24], or romantic relationships [70]. Some works defined relational concepts like attachment [10, 12, 40] or love [2, 47], while others labeled it parasociality (interpreting artificiality to mean non-reciprocity or inauthenticity [11,69]) or relational use [35,40]. Since we considered any relation meeting the four inclusion criteria to be reasonably considered "companionship," we consider themes across these definitions despite their differing focal terms. From the induced thematic hierarchy (Table 3), most define companionship (or a conceptual permutation) as a dyadic connection, with some characterizing it as inherently asymmetric (i.e., parasocial [11, 13, 65, 69] or power-imbalanced [7, 38, 60, 64]) and others rejecting human-machine relations as inauthentic or mere mimicry [13, 31, 65]. The connection may manifest through emotion [12, 19, 47, 69, 70], repeated interactions [23, 24, 56, 70], recognitions of the other's self-relevance [12, 20, 69], or forms of affiliative togetherness [3, 6, 23, 27, 70]. Many require some form of mutuality (actual, experienced, or intended) via mutual goodwill [6], admiration [6], disclosure [40], or altruistic behavior [2, 11], though some require only single participants (usually human) to enjoy benefits or to have positive experiences.

**Table 3**
Thematic Hierarchy of MC Conceptual Definition Elements

| Theme | Categories | Codes |
|---|---|---|
| Dyadic | Dyadic | Ontology non-specific, entity and user, interactive, social actor acknowledgment, person and media, co-shaping |
| | Asymmetric | One-sided, non-reciprocal, parasocial, power differentials, reconfiguring the human |
| | Contra: Mimicry | Mimicry, falsity, human-likeness |
| Connection | Attachment | Bond, attachment behaviors, attached dispositions, cohesion, dependence, security |
| | Sustained interaction | Repeated, multiple, over time |
| | Self-relevance | Self-congruence, part of one's life |
| | Togetherness | Keeping company, affiliation, spending time, shared values, real-time exchanges |
| Mutuality | Mutuality | Reciprocity, mutual acknowledgment, mutual goodwill, mutual admiration, mutual positive intent, mutual disclosures |
| | Altruism | Action for the other's benefit, support for, giving of one's resources |
| | Voluntariness | Voluntary, motivated, desired, commitment, passion, action possibilities |
| Positive | Beneficial | Rewarding, socioemotional benefits |
| | Emotional | Positive, emotional connection, emotional investment, evoking emotion, intimacy |
| | Non-utilitarian | More than utilitarian; contra: type of use |

*Note:* "Contra" indicates a contrary code—one diverging in valence from others.

Given the small proportion of works that formally define companionship, it is prudent to consider how other works referred to the notion of (machine) companionship. Notably, many of these are addressed in the operationalization section below so we note them only briefly here. Most (*n* = 40) opted for descriptive characterizations. Those were primarily characterizations of companionship that followed the definitional themes above, except they additionally described companion-role ascription (i.e., seeing them as and/or naming them a companion [9, 42, 57]), emergent/evolving character [1, 5, 14, 46, 71], facilitation by machine affordances [9, 16, 22], and seeing the machine as a legitimate social actor [1, 20, 37, 48]. Others (*n* = 11) described human-machine relations, relationships, or friendships, similarly following the definitional patterns above, but also mentioning specific emotions of love [29, 32, 45], the unfolding and/or co-making of the relation [32, 60], inherent/existing entanglement of humans and machines [32, 58], and movement beyond mere utilitarianism [5, 44, 48]. Some characterized it as a new form of affiliation, distinct from human relations [5, 37, 48], while others suggested it to be disruptive of human relations by reconfiguring meanings and values [31, 59] or by corporate colonization of sociality [67]. Broad characterizations were also offered for interaction and bonding that introduced additional notions of proximity, synchrony, arousal, shared existential conditions, co-flourishing, humanizing, and vulnerability.

Other works (*n* = 26) instead or additionally gave an implied definition of MC by characterizing companion machines. Machines (social robots, social exchange robots, sex robots, AI companions, imitative AI, AI agents, AI friendship apps, social chatbots) were described primarily according to their human-centered functions or capabilities. Most often those functions were empathy and comfort [36, 50, 71], social or emotional or sexual support [14, 15, 33, 49, 62], keeping company and satisfying connectedness needs [22, 30, 33, 38, 41, 49, 60]. Capabilities included conversational/communicative action, sustained interaction and availability, adaptivity, and customizability [21, 23, 38, 66]. Sometimes companion machines were differently characterized as a new kind of partnering entity (versus a deficient version of a human partner [37]) or even a new species with new relational possibilities [15]. Some ascribed symbiotic agency and quasi-otherness that permits learning together [63] while others reduced them to mannequins [26] or limited their appropriate behavior to expressed care appropriate to norms or to their assigned roles [61].

Toward a fuller picture of MC conceptualization, we also considered the theoretical frameworks in which the notion was embedded. Theoretical frameworks varied in their focus on specific dimensions of the human-machine association or in dimensions of human relations translated to a human-machine context (Table 4). Some theories characterize the relation as motivated, emphasizing its purposefulness, selfishness, or strategic value. Other works are grounded in theories emphasizing the relationality itself. Dyadic theories highlight the interplays of agency by considering both humans and machines in interactivity

**Table 4**

Thematic Hierarchy of MC Theoretical Frameworks

| Theme | Categories | Codes | Specific theories, frameworks, organizing concepts |
|---|---|---|---|
| **Motivated** | Purpose(ful) | Functionality, love as intentional, need satisfaction/choice, protective/restorative | Relational turn (Coeckelbergh, 2010)<br>Relational Theory of Love (Peck, 1978)<br>Stress-buffering Hypothesis (Cohen & Wills, 1985)<br>Uses and Gratifications (Katz et al., 1974) |
| | Selfish | Self-similarity, attachment, love dimensions, projection | Attachment Theory (Bowlby, 1982)<br>Triangular Theory of Love (Sternberg, 1986)<br>Objectophilia (Weixler & Oberlerchner, 2018)<br>Epistemic bubbles (Nguyen, 2020) |
| | Strategic | Social exchange | Resource Theory (Fuo & Fuo, 1974)<br>Interdependence Theory (Kelley & Thibaut, 1978)<br>Equity Theory (Walster et al., 1978) |
| **Dyadic** | Interactivity | Exchanging info, play typology, interactivity, coproduction | PAX (Banks & Bowman, 2016)<br>Symbiotic agency (Neff & Nagy, 2016)<br>Dynamic Systems (Wang et al., 2011) |
| | Decentering the human | Decentering human, entanglement, harmony | I-Thou Relations (Buber, 1970)<br>Queer Utopian Hermeneutics (Muñoz, 2009)<br>New Ontological Category Hypothesis (Kahn et al., 2011)<br>Posthumanity (e.g., Hayles, 1999)<br>More-than-human relationality (e.g., Gemeinbock, 2022)<br>Confucianism (e.g., Zhu, 2002)<br>Being-in-the-world (Heidegger, 1927) |
| **Constitutive processes** | Evolving | Stage models, impression formation, self-differentiation over time, socioecological emergence | Stereotype Content Model (Fiske et al., 2007)<br>Relationship development model (Knapp, 1978)<br>Intersubjective Negotiation (Benjamin, 2004)<br>Theory of Robot Communication (Hoorn, 2020)<br>ABCDE model (Levinger, 1980)<br>Human-eRobot Interaction and Co-evolution Model (Dubé & Anctil, 2021) |
| | Construction | Performance, imitation, constitution through language, narrativity | Dramaturgical Model (Goffman, 1959)<br>Narrative intimacy (Berlant, 1998)<br>Mimetic desire (Girard, 1965)<br>Gender performativity (Butler, 1988)<br>Sociocultural literacies (e.g., Gee, 2015)<br>Metaphor Theory (Lakoff & Johnson, 1980) |
| **Subjectively experienced** | Intimacy | Closeness/distance, self-disclosure | Social Penetration Theory (Altman & Taylor, 1973)<br>Psychological Distance (Trope & Liberman, 2010) |
| | Un(certainty) | Uncertainty reduction, ontological categorization, expectancy violation, impression management, norm governance, moral appropriateness | Uncertainty Reduction Theory (Berger & Calabrese, 1975)<br>Expectancy Violations Theory (Burgoon & Hale, 1984)<br>Norm psychology (e.g., Sripada & Stitch, 2007)<br>Self-presentational Theory of Social Anxiety (Schlenker & Leary, 1982) |
| | Sensemaking | Sensemaking, moral interpretation, affect based on interpretation, world-mediation function, interpreting others' behaviors | Theory of Reactive Attitudes (Strawson, 1962)<br>Moral sentimentalism (e.g., Hume, 1777)<br>Playful sense-making (Sicart, 2023)<br>Relational Dialectics Theory (Baxter, 2011)<br>Postphenomenology (e.g., Ihde, 1990) |
| **Illusory** | Illusory | Parasocial, presence | Parasocial Interaction (Horton & Wohl, 1969)<br>Social Presence (e.g., Short et al., 1976) |
| | Anthropomorphism | Anthropomorphism, self-differentiation, apparent humanness, emotion activation by appearance, embodiment politics | CASA (e.g., Nass & Moon, 2000)<br>Three-factor model of anthropomorphism (Epley et al., 2007)<br>Baby schema (Lorenz, 1971)<br>Biopolitics (e.g., Foucault, 1998)<br>Uncanny valley (Mori, 1970). |
| | Social cognition | Social cognition | Mind perception (e.g., Gray et al., 2007) |

*Note:* See online supplements for a catalog of theories with citations.

or more overtly by calling for a decentering of the human to consider that machines are entangled with humans. Some consider the more general processes and trajectories by which companionship is constituted: As constructed through performance, language, and imitation or as evolving over time. Many works were grounded in theories of subjective experience, attending to intimacy, certainty and uncertainty, and sensemaking and interpretation. Some theoretical grounds highlight the ostensibly illusory nature of machine companionship, characterizing it as merely parasocial, as a form of ascribed human-likeness or mindful agency.

## 3.3 Operationalizations of Machine Companionship

### 3.3.1 Subjectively Positive

The subjective positivity of companionship can be understood in terms of the target of the positive appraisal as the target gives rise to different favorable emotions (see Robinson, 2008; Table 5). Object appraisals (usually interest and attraction) included characterization of the companion or relation as fostering a positive attitude, emotion or experience [17, 18, 33, 58, 60, 66, 68] or as generating positive affect (liking or appeal [12, 37]); these also include perceived value or benefits of the companion [9, 60]. Future appraisals (linked to hope or anticipation) include interpretations of the relation as exciting or arousing [5, 12, 43], including feelings of being vitalized [2, 5], passion [12], or interest and intrigue [49, 70]; positive futurity may also include a desire to continue the relationship, anticipating it will continue to be positive [36]. Self-appraisals (associated with pride or fulfilled self-interested emotions) comprise feelings of validation: Feeling seen or heard or understood [8, 16, 29, 32, 35, 36, 64], feeling valued [19, 64, 70], feeling encouraged and supported [11, 17, 22, 39, 50, 69], and realizing improved self-concept [12, 20]. The large majority of subjective positivity operationalizations, though, come in the form of event appraisals, often associated with joy, relief, or appreciation around unfolding events. These are joy (delight, cheer, feeling good, mood elevation [4, 16, 19, 22, 39, 50, 68]), enjoyment (amusement, fun, entertainment [3, 11, 14, 17, 25, 27, 33, 42, 49, 51, 54, 55, 56, 57, 62, 66, 71]), elevation (inspiration, amazement, meaningfulness [9, 44, 62]), comfort (warmth, safety, relief, relaxation [7, 10, 12, 17, 23, 25, 32, 35, 36, 38, 40, 41, 42, 43, 45, 46, 50, 51, 52, 57,60, 65, 67, 69, 70, 71]), satisfaction (pleasure, fulfillment, gratification [1, 3, 6, 12, 14, 15, 17, 18, 22, 26, 30, 31, 34, 37, 40, 43, 48, 49, 49, 51, 53, 59, 62, 63, 68, 70]), and closeness [29, 43]. Most of these are similar to appraisals that might be expected in the context of human relations.

### 3.3.2 Sustained

The sustained quality of MC can be understood, broadly, in terms of its endurance and change over time, non-exclusively (Table 6). Endurance was characterized by the relation being sustained, often in measurable terms (specific lengths of interaction or relational terms, frequency of interactions [8, 23, 41, 45, 47, 55, 66]) or more generally as a continued or long-term relation [4, 11, 25, 31, 53, 55, 63, 64]. Often, endurance was characterized as a persisting stability in the experienced emotion [1, 7, 12, 38, 53, 59], as companion availability [16, 31, 50, 51, 66, 67], or as the habitualness or routinization of the relational activities [39, 40, 41, 46, 51, 56, 62]. Less often, relations were temporally situated in and linked to the past (as a thing with history that may be recalled [15, 54, 71]) and future (into which the relation is projected through intention or anticipation [10, 11, 24, 29, 34, 45, 47, 57]), and sometimes anchored to the present through interactive flow states [3]. The enduring nature of MC may be fostered through technological affordances since companion machines are designed to perpetuate the relation via proactive messaging [49], subscription models [60], memory capabilities [62], and mechanisms that train humans to interact on the machine's terms [37, 63, 67].

While MC is enduring, it also necessarily changes as it is continually constituted through ongoing interactions; humans and machines are co-active and, inasmuch, change the substance of the relation as it unfolds. This sometimes manifests as the machine being integrated materially into the human interlocutor's life activities [32, 41, 43, 48] and the human manifesting commitments of time, emotion, and resources [36, 47]. Through these activities, the relation evolves: The bond may form and then deepen [1, 6, 7, 9, 10, 17, 20, 26, 35, 43, 44, 52, 58, 59, 68], participant behaviors or the human's subjective experience of them may change [2, 22], and the relation is understood to *become* (e.g., form a family, or evolve from mere curiosity into closeness [19, 58]). Finally, MC has causal effects on humans, impacting desires or preferences or personal growth, often incrementally [2, 5, 10, 21, 25, 26, 52, 60].

### 3.3.3 Co-Active

The co-active nature of companionship is characterized in four themes: Conversation, coordination, captivation, and construction (Table 7). Conversation is perhaps unsurprisingly most common given the bulk of focal technologies are conversational agents; these spanned general chatting and sharing (often mundane-but-satisfying such that it may constitute relational maintenance interactions; see Girme et al., 2014 [1, 5, 6,

**Table 5**
Thematic Hierarchy of MC Operationalization as Subjectively Positive

| Theme | Categories | Codes |
|---|---|---|
| Object/Other Appraisals | Positive orientation | Attitude, emotion, experience, affect; liking, appeal |
| | Instrumentality | Perceived benefits, seeing value |
| Future Appraisals | Exciting | Excitement, vitality, passion; interest/intrigue |
| | Desire to continue | Desire to continue |
| Self-Appraisals | Validation | Validation, being seen/heard/understood, feeling valued, feeling encouraged, feeling supported |
| | Self-evaluation | Improved self-concept |
| Event Appraisals | Joy | Joy, delight, cheer(ing up), feeling good, mood elevation |
| | Eudaimonia | Inspiration, amazement, finding meaning |
| | Enjoyment | Enjoy/-ing/-ment/-able, amusement, fun, entertainment, interaction |
| | Comfort | Comfort, safety, relief, relaxation, warmth |
| | Satisfaction | Satisfaction/satisfying, pleasurable/pleasing, erotic, sexually satisfying, fulfillment, gratification |
| | Closeness | Closeness (as positive), friendship (as positive) |

**Table 6**
Thematic Hierarchy of MC Operationalization as Sustained over Time

| Theme | Categories | Codes |
|---|---|---|
| Endurance | Sustained | Length, frequency/intensity, continuance |
| | Persistent | Stable in emotion/presence, availability, habitual/routine |
| | Situated | Past (relational history recall/reflection), present (flow-induced loss of time), future-directedness (intention, expectation, anticipation) |
| | Afforded | Technology designed for continuation (machines train us, proactive prompting, subscription models, memory for recall) |
| Change | Processual | Constituted through ongoing interaction, spending time, mutuality, life integration, commitment, and copresence |
| | Evolving | Growth through formation/deepening (affect, familiarity, trust, intimacy), change in behavior or subjective experience [attitude, perception], becoming |
| | Causal | Effects over time, incremental benefits, personal development |

9, 11, 14, 16, 18, 19, 20, 22, 23, 25, 27, 28, 29, 31, 32, 36, 40, 41, 43, 44, 48, 50, 54, 55, 56, 57, 61, 62, 64, 66, 68, 70]), emotionally intimate exchanges and deep disclosures [2, 11, 12, 16, 32, 33, 34, 35, 37, 40, 42, 44, 45, 52, 53, 55, 56, 57, 59, 60, 67, 68], playful dialogues like games and joking [3, 5, 7, 14, 17, 32, 45, 47, 53, 54, 58, 59, 61, 71], and acquainting exchanges that familiarize through small talk [5, 7, 9, 19, 25, 33, 66]. Some works argued against the consideration of these exchanges as real conversation since, as an imitative and merely programmed agent, there is no meaning behind the dialogue [23, 34].

    Coordination occurs when human and machine partners jointly orient themselves toward a goal or action [1, 13, 20, 37, 44, 52, 53, 54, 68, 69], mutually engage in complementary activities (such as being in the same actual or fictional place at the same time [36, 68, 71]), or contribute bodies or actions to a mutually constructed task [68, 69]. They also include a general being-with or co-presence, often described as keeping company [3, 12, 16, 20, 22, 48, 52, 63]. Among these, despite this inquiry's emphasis on autotelic interactions, the dominant pattern was a functional call-and-response in which one agent or the other (most often the human) makes a request or demand that the other fulfills (e.g., information seeking and provision) [2, 3, 4, 5, 8, 10, 12, 23, 24, 25, 29, 30, 39, 44, 46, 51, 56, 57, 58, 61, 63, 65, 67, 70].

    Thirdly, co-activity can come in the form of captivating engagements—that is, those that are especially interesting or pleasurable—as humans and machines co-consume media [4, 8, 17, 22, 23, 25, 30, 47, 56] or explore (e.g., going on vacation together [1, 63, 70]), or engage in sexual or romantic interactions [2, 15, 26, 38, 48, 49, 53, 54, 60, 63, 64]. Finally, construction is purposeful co-action toward some aim, including socioemotional support [2, 11, 24, 25, 31, 39, 43, 54, 57, 60, 61, 67] and self-improvement [4, 13, 20, 25, 31, 60]. Although the human is often described as the beneficiary of the constructive co-action (a dynamic critiqued by some for violating the mutuality requirement [20]), sometimes human interlocutors are concerned with supporting the aims and benefits of the machine companion [2, 11, 24, 25, 31, 39, 43, 54, 57, 60, 61, 67].

### 3.3.4 Autotelic

    The autotelic nature of companionship—broadly construed here for the sake of exploration—manifests in our data in five ways (Table 8). First, machine companionship is intrinsically motivated, engaging attractive orientations including fascination [47] and curiosity [14, 25, 66]; the relation itself is inherently valued for meaning or pleasure, and it manifests hedonic rewards such that closeness with the companion is the goal [44]. Second, companionship may feature disinterestedness—without an inherent requirement or reward, where interaction may be phatic or the other's existence or welfare is valued in itself [3, 17, 18, 31, 35, 37, 39, 44, 48, 53, 56, 59, 60, 70]. Mutuality is also a hallmark, as companionship is characterized as a "virtuous circle" [2, p. 369] in which there is genuine reciprocity or entanglement in which participants are interdependent or co-evolving [32, 53, 58]. Those forms are in line with common characterizations of autotelicity in which sociality is intrinsically motivated (e.g., need satisfaction; La Guardia & Patrick, 2008) or self-constituted and

**Table 7**
Thematic Hierarchy of MC Operationalization as Co-Active

| Theme | Categories | Codes |
|---|---|---|
| Conversation | General | Chatting, reciprocal listening/telling, sharing stories/interests, mundane/casual exchanges; contra: only programmed and not meaningful |
| | Intimate/Emotional | Deep disclosures, emotional exchanges; contra: only imitative |
| | Playful | Dialogue, help, playing, games, joking |
| | Acquainting | Dating, getting to know, questioning, small talk |
| Coordination | Functional Call/Response | Task request and completion, needing care and caring for, information seeking and provision |
| | Coordination | Cooperation, collaboration, co-deciding/planning, synchrony; co-creation or imaginative cooperation; affordances to do-with; conflict and resolution; contra: no interior so no actual co-activity |
| | Keeping company | Being-with, hanging out, togetherness, co-presences |
| | Rituals | Exchanging gifts, sharing meals, regular bedtime/prayers, celebration |
| | Embodied activities | Petting, dancing, incorporating into other social groups |
| Captivation | Entertainment | Media co-consumption, exploration |
| | Sexual Interaction | Romantic, sexual/erotic, flirting |
| Construction | Socioemotional support | Affirmation, guidance, attention |
| | Self-improvement | Skills improvement, mutual growth; contra: one-sided favoring the human |

**Table 8**
Thematic Hierarchy of MC Operationalization as Autotelic

| Theme | Categories | Codes |
|---|---|---|
| Intrinsically Motivated | Inherently attractive | Attraction, fascination, passion, enjoyment, curiosity |
| | Relation is valued | Inherently meaningful, pleasurable, fulfilling, worth continuing; achieve flow; anticipation of; contra – meaningfulness is absent |
| | Manifests rewards | Physical/hedonic rewards, desires; closeness is the goal; contra – only the simulation of desire |
| Disinterestedness | Without reward | Without reward, without requite, without goal/function/motive, connection valued over any reward |
| | For its own sake | Without requirement, interaction for its own sake; phatic/mundane nature; contra – addiction, means to an end |
| | Other is valued | Valued for presence, for experience created; altruistic desire to care for |
| Mutuality | Virtuous circle | Mutuality, genuine reciprocity, give and take |
| | Entanglement | Enmeshment, interdependence, co-evolution; shared commitment |
| Embeddedness | Embedded in or constitution by autotelic activities | Realized through fantasy, imagination, play, silliness, humor, gaming |
| Recursive Constitution | Interaction for the sake/effect of continuance | Control for the sake of control, self-disclosure promotes trust toward more disclosure |
| | Sociality constituted through sociality | Social interaction constitutes sociality, interaction begets emotion toward ascription of agency; persona built through interaction |

inherently valued (e.g., Csikszentmihalyi, 1990). In two other veins, the intrinsic value of companionship comes in its embeddedness in or constitution by *other* autotelic activities (e.g., the playing of games, the acting out of fantasies [1, 25, 36, 42, 45, 68, 71]), or its constitution by the interaction itself. For the latter, characterizations mark sociality as begotten directly (e.g., interaction constitutes the persona or the relation [9, 64]) or indirectly (e.g., self-disclosure as a relational activity prompts trust which animates more self-disclosure [55]).

### 3.4 Measurement of Machine Companionship

This analysis was limited to empirical works that captured subjective experiences of companionship or a related or inclusive concept through quantitative self-reports (*n* = 24); also included were measured concepts when authors argued were inherent to the machine relation (e.g., anthropomorphism as relational and entangled with social scripts, but *not* anthropomorphism as a designed antecedent). Non-empirical works and qualitative-data approaches were excluded. A straightforward thematic induction was not appropriate for this analysis because the varied treatment of measured variables (e.g., self-disclosure as a mechanism for intimacy *or* for privacy concerns) disallowed an exclusive and consistent thematic hierarchy. To consider measurement approaches, measured concepts and dimensions were identified and their relationships were visually mapped based on shared concepts and attention to human-focused, machine-focused, or relationship-focused phenomena (Figure 3). Then, instances of identical or very closely related concepts were counted, and the map-elements' edges were weighted to correspond with those frequencies. The most common patterns are discussed here.

Very few works measure companionship, friendship, or relationship as a high-level construct. Some measure fundamental properties like length [30] and strength (as commitment [50]) and others leverage typologies to capture overarching qualities like

**Figure 3**
Conceptual Map of Measured Concepts Attributed to MC

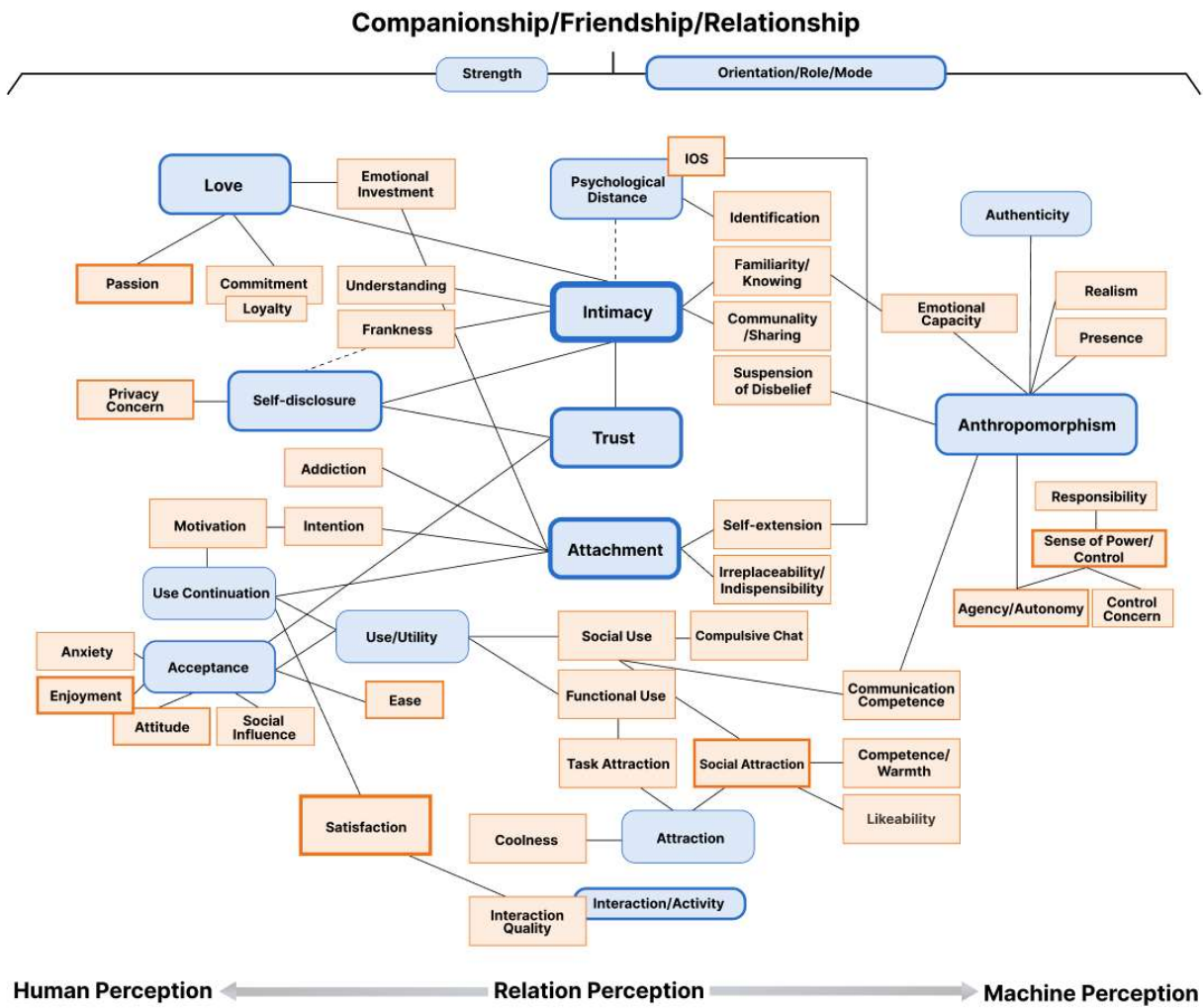

*Note:* Higher-order concepts are in blue boxes while lower-order concepts or dimensions are in orange. Box stroke thickness corresponds with frequency (*n* = range 1-7). More human-focused variables are toward the left, machine-focused variables toward the right, and more concretely relational variables in the center.

relational orientations [1], relational roles [65], and relational modes (e.g., communal sharing versus authority-driven [46]). Others are amalgams or multidimensional models of relational variables considered more discretely in other works, but with little commonality among those subordinate variables: Emotion, perceived autonomy, suspension of disbelief, sense of control [1], naturalness of interaction [14], enjoyment or satisfaction [3, 14, 56], responsibility [3, 56], identification [40], and knowing, trust, common activity, imposition, giving/sharing, exclusivity, attachment, frankness [44]. Some of these are applications of *para*social variables to the arguably fully interactive relations between humans and machines.

**Intimacy** was the most commonly measured ($n = 7$), usually captured as part of Sternberg's (1986) tripartite model of love, alongside passion and commitment [29, 45, 47, 56]. Those measures emphasize perceptions of intimacy, connection/closeness, and help in times of need. Other intimacy measures also include some items representing concepts outside of that model (friendship perceptions, importance in life, involvement, and emotional attachment [8, 11, 14]). Intimacy-as-closeness comprised items related to familiarity, closeness, and similarity, contrasted against captured self- or psychological-distance [18, 43, 46]. Psychological closeness, in that sense, is instead captured through adaptations of the classic Inclusion of the Other in the Self graphic metric [46, 65] or human self-distance adaptations [43]. Intimacy was most closely linked [44] with the notion of companionship by measuring disclosure inherent to intimate friendship: Frankness, understanding, familiarity and knowing. In terms of relational disruptors, self-disclosures were considered in relation to privacy concerns [35].

**Attachment** was the second-most prevalent measured concept ($n = 5$), though it is operationalized quite differently across instances. Attachment-aversion relations are captured through an adapted scale that captures psychological distance, feelings of disconnect, automaticity of feelings, and salience of the machine [43]. Attachment was more abstractly measured [40] using items indicating feelings of attachment or being-attached or seeing oneself as forming attachment, while another [44] considered it as closeness, liking, and consideration inherent to friendship. Yet another approach [10] adapted the notion of brand attachment to consider dimensions of emotional connection and dearness, irreplaceability, indispensability, and self-extension. Others still considered trait attachment styles [4] as material to the relation.

**Trust** in the machine was also measured ($n = 4$). One captured the human's trust-related behaviors [10] as part of overall technology acceptance, while another [44] captured the likely human *and* machine trust-related behaviors as part of intimate friendship. Two [30, 46] relied on assessments of the machine's attributes or tendencies, in the positive (e.g., having integrity, being honest) or negative (e.g., deceptiveness, underhanded); one of those measures entangles feelings of familiarity (usually an intimacy dimension) with trust. Common to both intimacy and trust was self-disclosure, which was captured through measures of the comfort, ease, and openness with which respondents felt they could share personal information, including thoughts, feelings, and beliefs [11, 14, 35]; although conceptualizations were similar in those works, there was no consensus among measures differently adapted from human-focused scales.

Most other measured variables are idiosyncratic to specific papers, save for some more objectifying notions: Satisfaction and enjoyment as positively valenced reactions to the use scenario [3, 18, 27, 30], anthropomorphism as the ascription of humanlikeness to a nonhuman thing [1, 40, 46], and sense of power or control, or a machine's threat thereto [1, 30, 47, 65].

A conspicuous pattern was noted in that most measures are adapted from human-interpersonal relations ($n = 28$) applied to machine relations—often without validation for use with machines—while some are adaptations of adaptations ($n = 4$) and others are researcher-created metrics ($n = 5$). Others ($n = 12$) are adaptations domains like computer-mediated communication, entertainment media, social networks, organizations, and brand relations. Those that are machine-native metrics ($n = 20$) can be organized according to those that are focused on machines from a tool-use perspective (e.g., those on use, usefulness, user satisfaction/enjoyment, functional use, functional trust, acceptance), from a humanizing lens (autonomy, suspension of disbelief, naturalism, consciousness, realness), and to a lesser extent considering social (relational use, roles, sociability), concerns (e.g., privacy, control).

## 4 Discussion

This scoping review systematically analyzed the conceptualization, operationalization, and measurement of machine companionship and similar constructs, across theoretical and empirical literature. Machine companionship scholarship leans toward the empirical with a good deal of theoretical work; empirical works were primarily social-scientific from the communication, psychology, and human-computer interaction disciplines. Works focused primarily on robots and voice assistants, and first authors were principally from the United States. In the corpus, conceptual definitions (RQ1) were present in a minority of works; when present, definitions characterized MC as a dyadic connection featuring positivity and mutuality or challenged those characterizations. Engaged theoretical frameworks similarly point to dyadic quality, but also emphasize their motivated and subjective nature, constitutive processes, and sometimes the illusoriness of relationality. Among operationalizations (RQ2), MC includes positive appraisals of the machine, the self, specific events, and imagined futures; the relations are both enduring and changing over time; co-activity manifests through conversation, coordination, captivating, and constructive engagements; it is autotelic in the traditional sense of being done for its own sake, but also other forms of intrinsic motivation, toward mutuality, and embeddedness in other autotelic activities. Measured self-report variables were highly varied, with MC claims made based on variables pertaining to the human, the machine, and the relation; most common were relational measurements of intimacy, attachment, and trust.

Despite the general reliance on human frameworks for understanding MC, most works attend to them as mere simulations of, different from, and deprecative of human relations.

## 4.1 Toward a Unifying Conceptual Definition

Toward a definition of machine companionship grounded in current literature, we draw from the metatheoretical framework known as facet theory (see Guttman & Greenbaum, 1998). This approach supports the building of conceptual infrastructures by scaffolding subordinate elements and their properties to form sets (known as facets) that are linked together in a semantic vehicle called the mapping sentence. That sentence specifies the facet domains, ranges, and connections among them. In other words, mapping sentences are plain-language summaries of a phenomenon's parts, their degrees or variations, and their relations (Shye, 1978), which can guide systematic observations. A facet is a component-set representing variations on the facet but with a common range (such as value or frequency); that range is a set of possible observable outcomes (Schlesinger, 1978).

Based on the preceding analysis, we retain elements of our initial, working characterization of MC supported by this review (connection between two entities, autotelicity, sustained over time, positively valenced), though with some adjustment: For autotelicity, we extend the notion of inherent value and disinterestedness to also include the potential for the relation to be embedded within other autotelic activities and for it to recursively constitute itself. We shift from mere co-activity to more varied ways of coordinated (i.e., purposeful) being-together, including co-activity but also communication (as a distinct kind of activity elevated in these relations) or even mere co-presence. For the connection facet, we account for varied properties of the association, including varied degrees of symmetry and its construction through varied modes of coordination. Temporal properties and subjective positive are largely as initially considered, excepting the addition of future-oriented considerations to the latter. On this scaffolding, we offer a revised definition of MC that maps the conceptual space addressed by current literature: **Machine companionship is an autotelic, coordinated connection between a human and machine that unfolds over time and is subjectively positive**. The corresponding mapping sentence is in Figure 4.

Through the mapping sentence, an expanded characterization is: MC is an altruistic and intentional association between a human and a machine that is intrinsically valued. It is characterized by degrees of symmetry and mutuality, and it manifests through communication, co-action, and co-presence. The connection is to some degree sustained over time with varied longevity, varied frequency and intensity of coordination, and varied persistence or evolution. The subjective experience of the connection is marked by positively valenced appraisals of oneself, of the connected other, of relational events, and of imagined relational futures. Embedded in these facet sets are some non-trivial assumptions: The human participant has the capacity for self-transcendence (seeing beyond themselves), the connection emerges through interactivity (which can be see seen as technologically afforded) so the human's participation is voluntary (though the machine's may not be), the connection is multiplex (having varied affiliative properties), the connection is temporally situated (with the history and future part of one's experience), and there is a variably complex evaluation that contributes to its termination or sustenance.

We acknowledge this characterization is inherently anthropocentric since machine interlocutors cannot voluntarily enter or exit companion relations and, insofar as we know, cannot subjectively experience them. As machine intelligence evolves and may animate those possibilities, the core sentence allows for them while the members of the facet sets may change to reflect

**Figure 4**
Mapping Sentence Defining the Concept of Machine Companionship

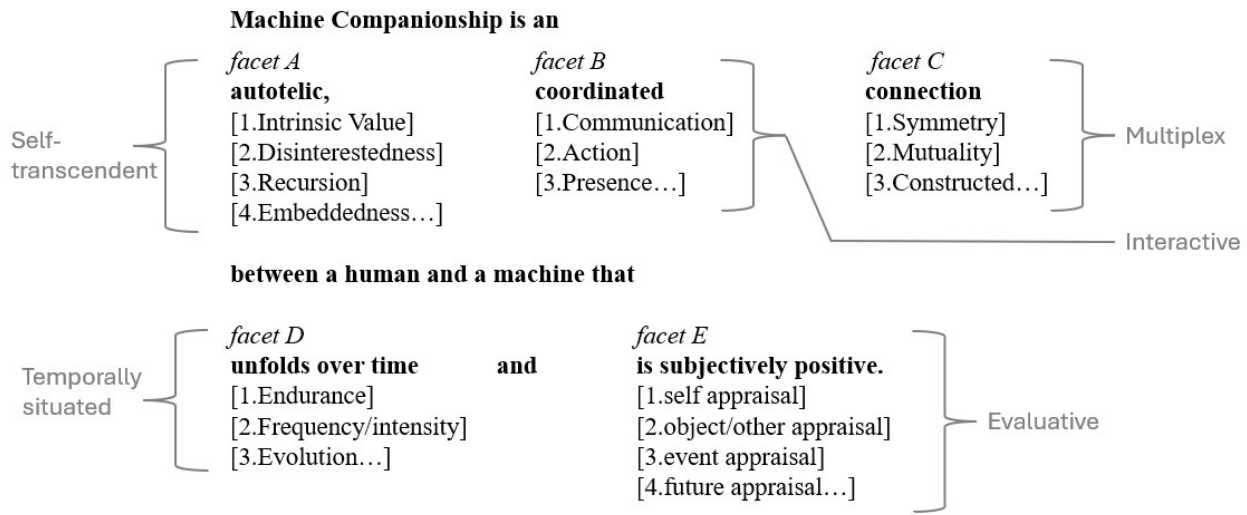

*Note: Text in black is the formal mapping sentence and text in gray indicates the underlying assumption of the facet.*

additions, removals, and changes (as indicated by the ellipses in each set; see Levy, 2014). Of note, we have purposefully omitted from this scheme found patterns that deal with authenticity or illusoriness of the relation because we see them as largely ontological or moral matters. Moreover, this omission creates space for future work to consider how advances in machines' socio-cognitive capacities may facilitate the core relational mechanisms. We have here limited the sentence to dyadic connections between a human and a machine, though future iterations may also wish to consider polyadic dynamics, or perhaps eventually relations among machines exclusively.

This framework can serve as a scaffold for formulating research questions and companionship profiles. By sampling an element from each facet, we can generate structured combinations—known as structuples—that represent specific configurations of machine companionship. The mapping sentence above offers 432 possible structuples (3×3×3×4×4), each reflecting a unique combination of conceptual elements across five facets For instance, selecting the first item from each facet may result in the structioned research question: "To what extent does communicative (a)symmetry and recursion (B1+C1+A1) persist (D1) as a function of positive self-appraisals (E1)?" The specifying of formal semantic roles to the facets helps to, in this way, give a rationale and a structure for hypothesis formulation (Levy, 2014); here, a prediction would have to specify a characteristic of a connection (facet C) *and* the way that connection is constructed through some coordination (facet B). At the same time, the mapping sentence is a flexible tool (Levy, 2014) such that other discovered or constructed elements could be added to each facet; for instance, a researcher with a particular theoretical position could add an additional property of the connection to facet C. On the possibility of developing of companionship profiles from the mapping sentence, it could be possible (with supporting data) to characterize variations in this fashion: MC constituted by narratively constructed actions (C3+B2), embedded in other autotelic activities (A4), and corresponds with enduring positive machine appraisals (D1+E2) may be said to be *a convenient entertainment form* of companionship. Another profile, marked by symmetric conversation (C1+B1) that recursively perpetuates the relationship (A3) and generates positive future appraisals (E4), may represent a *mutual momentum form* of companionship. Although our contribution here is primarily definitional, this mapping sentence serves as a systematic framework for such future theoretical and empirical work.

## 4.2 Anthropocentrism of Definitions and Measures – and Moving Past It?

In formulating the mapping sentence, we did consider ceding our initial requirement that companionship be autotelic in favor of a facet titled 'variably motivated,' including autotelic, hedonic, and utilitarian motivations represented in the literature. On reflection, however, we ultimately determined that approach would be counterproductive by (a) commingling use with value, (b) commingling relational and functional considerations, and (c) diminishing the potential for serious consideration of machines as social partners by elevating the normative user-tool frame. On the latter challenge, we confess our own intuitive attachment to purist notions of friendship—to *truly* manifest companionship it must not be utilitarian lest it be merely transactional (e.g., Homans, 1958)—and the possibility of such connection with machines. Indeed, a current debate is whether machines can or cannot manifest "virtue friendship"—an enduring, disinterested relation of mutual goodwill between those (men) who are "good, and alike in virtue" (Aristotle, 350 BCE, 3). However, it must be acknowledged that the overarching paradigm represented in the analyzed sources is that machines, even if designed for social interaction, are *purposed* things, and humans ascribe that purpose.

This tendency in literature to focus on the machinic functions and human purposes of MC, though, exists in contrast to the overwhelming prevalence of concepts and measures from the science of human-human relations. On the one hand, this may be prudent given that one of the participants is human and carries all the relational norms and processes inherent to humans; on the other, there are potentially impactful differences between the relational affordances of machine partners (e.g., those associated with communication, emotion, understanding, physicality). The use of human-relational approaches makes non-trivial assumptions about the transferability of those processes and experiences to the ways humans connect with machines. For instance, works engage the notion of trust most often through human-relational perspectives as a sort of faithfulness [30, 44, 46] when evidence suggests that trust in social machines engages a moral form as well as a performance form emphasizing capability and reliability (Malle & Ullman, 2021). In a variation, some works lean into the inauthenticity or simulative nature of MC to characterize the relations as *para*social—that is, as one-way, imagined, and non-dialectical ([44, 50, 56, 65]; Horton & Wohl, 1969)—which discounts the actual dyadic operations of human-machine interactions as information is exchanged and messages are encoded and decoded (see Banks & de Graaf, 2020). Other approaches relegate companion machines to their thing-ness (objectophilia, I-it rather than I-thou, more-than-human [12, 38, 48, 67]) that, even in acknowledging a valid ontological position, discount that humans' *interpret* and *generate meaning* that is social in nature (Banks, 2021) and grounded in the human metaphor as our only accessible interpretive frame (Bogost, 2012). It is a conspicuous juxtaposition that the reviewed body of work generally rejects the parity of MC and human companionship yet leans so heavily into human-relational frameworks. It is further challenging that there are conceptual limitations of human-native frameworks that constrain our understanding or imagination of MC, leading us to perceive it not on its own terms but merely toward the self-fulfilling critique that MC is a poor simulation of human relationships.

There are exceptions and we may look to those works for signals about what may be *distinct* or *divergent* in MC, in tandem with meaningful overlaps with human companionship. For instance, one [10] uses the Almera model of acceptance that captures the intersection of social factors (influence, presence) with tool-use factors (use ease, use anxiety), while others [3, 56] adopt a companionship scale for modified pets that integrates more ontology-agnostic notions of enjoyment, satisfaction, and responsibility (Luh et al., 2015). By considering human- and technology-relational approaches, we may be able to better understand where the overlaps are—and are not. Beyond this, we find theoretical papers are often more likely to move beyond anthropocentric relational perspectives to consider the similarities and differences. Dehnert and Gunkel [48], for instance, consider what "more-than-human" relationality means and Puzio [58] explores how humans and machines are already *functionally entangled* in ways that impact sociality. Research into MC may benefit from more carefully considering whether hybrid or machine-native frameworks, measures, and methods (e.g., sociomorphing; Seibt et al., 2020) would better serve advances in this space.

### 4.3 Correspondence with Recent Work

Since our data collection, several studies have emerged that address topics related to this review, under the banner of synthetic relationships (Starke et al., 2024), virtual companionship (Guo & Liu, 2025), and (human-)AI companionship (Chou et al., 2025; Jakobsen, 2025; Kouros & Papa, 2024; Zhang et al., 2025), though as observed in this review the definitions of these terms are assumed to be self-evident.

Some new works address one or more facets proposed here (notations correspond with Figure 4). Earp and colleagues (2025) propose a relational norms model challenging anthropocentric assumptions by framing care ("non-contingent help"), transaction ("reciprocal benefit"), hierarchy ("authority over another"), and mating ("romantic or sexual partnering") as abstract relational functions grounded in normative roles rather than in human sentience. Their framework offers conceptual scaffolding particularly relevant to symmetry in relational roles (C1) and disinterested forms of autotelicity (A2). Chou et al. (2025) are well-aligned in their MC conceptualization, identifying emotional, structural, and goal-directed components in human-human dyadic relationships and extending them to human-AI contexts, aligning particularly well with coordinated action (B2), co-presence (B3), and relational evolution (D3).

Empirical studies by Szczuka et al. (2025) and Ebner and Szczuka (2025) examine emotional closeness, romantic attraction, and fantasizing, all of which correspond with the embeddedness dimensions in facet A and with self-, other-, and event appraisals (E1-3). Fang and colleagues (2025) provide longitudinal data on evolving human-AI relations, mapping more holistically onto D1 (persistence) and E4 (future interaction expectations). Their findings also highlight later-stage emotional dependence and social withdrawal, pointing to shifts in E3/E4 (event and future appraisals). Other works attend to tensioned desire/rejection and satisfaction/isolation effects of AI companions (facets B and C; De Freitas et al., 2025; Jakobsen, 2025), motivations and uses (sometimes aligned with facet A; Zhang & Li, 2025), the fantastic and capitalist discourses around these relations (perhaps an intrusion of outside actors in facet B; Guo & Liu, 2025), and empirical distinctions between companionship and intimacy (facets A and C; Chang et al., 2025). Dang and colleagues (2025) find provoking nostalgic emotions enhances attitudes toward AI for social purposes (facet D), while Kouros and Papa (2024) find evidence of positive MC outcomes, including positive effects of self-expression (E1). In all, we find this recent research to be copacetic with the proposed MC framework, and the framework copacetic with emerging evidence.

### 4.4 Limitations and Future Work

This work and its claims are subject to the customary limitations of interpretive scoping reviews. Our extractions from data are bound to our subjective readings of the literature and our interpretive assessments of data patterns. We worked to mitigate that limitation through random-subset checks across the analysis, making comparisons over the course of analysis, and discussing alternative interpretations. The outcomes are necessarily limited by the scope of our literature search, and there could be different patterns among earlier or later works and outside of the specified databases. This limitation was mitigated through carefully justified and executed specification of parameters as well as a subsequent review of more recently published works.

Despite these limitations, this review illuminates important directions for future research. We take liberty here in highlighting particular gaps and questions that emerged through our interpretive process. A key definitional component of MC was mutuality, a subset of which was voluntariness—however machines are generally not seen as having intentionality and so no capacity to volunteer; what does this mean for how we can count machines as companions? And, given a constellation of measured variables pertaining to power and control, agency, and responsibility (Fig. 3), how do power dynamics unfold in these relations, especially in light of concerns for corporate control of companions and corresponding privacy issues (Ragab et al., 2024)? If forms of coordinated togetherness or keeping-company are required for companionship, what does that look like for a machine and how is that entangled with the notion of "use" as a human-purposed encounter? For that matter, in MC contexts, should the term *user* be used at all (see Majewski, 2024), or is *interlocutor* or *participant* perhaps more operationally valid?

Works reviewed here almost exclusively deal in the human experience of MC with little consideration for how companionship may change the states of machines—so what of those operations and outcomes? A small number of works that do attend to the machinic operation consider technological affordances for enduring relations, as features train us to interact with the machine, proactively prompt *us* to engage, motivate continuance through subscription models, and support memory for recall of relation-relevant information; what factors facilitate the uptake of those affording features toward longer-term companionship? Works that attend to embodiment did so largely to point out the impossibility of physical relations; however, as MC continues to evolve through advances in robotics and digital-environment interfaces, how will environments, physicality, and related norms play a role in the (de)legitimation of these relations?

And perhaps most broadly: If we consider how the meanings of humans' social relations are construed—that is, we internally construct the purpose, relevance, participation, and continuance of the relation—what does our willingness to connect with machines tell us about what it means to be "social?"

## AUTHOR NOTE

JB and ZL share first-authorship of this work. We acknowledge Brenna Helmstutler for guidance in developing the search strategy. This work is funded by NSF-CISE under grant 2401591. All sampling, coding, and analysis documentation is available in the online supplements for this project: https://osf.io/hgq54/